\documentclass[10pt]{article}

\title{Evolution of entanglement for spin-flip dynamics}
\author{Slawomir Koziel \\
Institute of Experimental Physics, Gdansk University, \\
Wita Stwosza 57, 80-952 Gdansk, Poland \\ \emph{e-mail:} koziel@iftia6.univ.gda.pl \\
Wladyslaw A. Majewski \\
Institute of Theoretical Physics and Astrophysics, Gdansk University, \\
Wita Stwosza 57, 80-952 Gdansk, Poland \\ \emph{e-mail:} fizwam@univ.gda.pl }

\date{}

\newtheorem{deff}{Definition}[section]
\newtheorem{lem}{Lemma}[section]

\newtheorem{uw}{Remark}[section]
\newtheorem{stw}{Proposition}[section]

\newcommand{\dw}{\emph{Proof.$\;$}}
\newcommand{\finito}{}%{\begin{flushright} $\Box$ \end{flushright}}
\newcommand{\beq}{\begin{equation}}
\newcommand{\eeq}{\end{equation}}
\newcommand{\beqar}{\begin{eqnarray}}
\newcommand{\eeqar}{\end{eqnarray}}
\newcommand{\Cn}{{\setbox0=\hbox{
$\displaystyle\rm C$}\hbox{\hbox
to0pt{\kern0.6\wd0\vrule height0.9\ht0\hss}\box0}}} %Complex
\newcommand{\hj}{{\mathcal{H}}_1}
\newcommand{\hd}{{\mathcal{H}}_2}
\newcommand{\oti}{\otimes}
\newcommand{\hjd}{\hj\oti\hd}
\newcommand{\si}{_{i=1}^n}
\newcommand{\sj}{_{j=1}^m}

\newcommand{\la}{\langle}
\newcommand{\ra}{\rangle}
\newcommand{\bla}{\big\langle}
\newcommand{\bra}{\big\rangle}

\newcommand{\ovl}{\overline}
\newcommand{\udl}{\underline}
\newcommand{\prw}{^\frac{1}{2}}
\newcommand{\mprw}{^{-\frac{1}{2}}}
\newcommand{\fal}{\widetilde}

\newcommand{\xjj}{\xi_1\oti\xi_1}
\newcommand{\xjd}{\xi_1\oti\xi_2}
\newcommand{\xdj}{\xi_2\oti\xi_1}
\newcommand{\xdd}{\xi_2\oti\xi_2}
\newcommand{\nonu}{\nonumber}
\newcommand{\iden}{{\bf 1}}
\newcommand{\rzjj}{|\xi_1\ra\la\xi_1|}
\newcommand{\rzjd}{|\xi_1\ra\la\xi_2|}
\newcommand{\rzdj}{|\xi_2\ra\la\xi_1|}
\newcommand{\rzdd}{|\xi_2\ra\la\xi_2|}

\newcommand{\trjr}{Tr_1(\rho)}

\newcommand{\bazajd}{\{\varphi_i\oti\phi_j\}}
\newcommand{\bazaj}{\{\varphi_i\}}
\newcommand{\bazad}{\{\phi_j\}}
\newcommand{\fsh}{F\oti id_{\hd}}
\newcommand{\gsh}{id_{\hj}\oti G}
\newcommand{\fr}{f=\fsh}
\newcommand{\gr}{g=\gsh}
\newcommand{\korel}{\bla E(f)g\bra_{\rho}}
\newcommand{\rozred}{\fal{\rho}=\sum_i\lambda_i\rho_i^{II}}
\newcommand{\rozredf}{\fal{\rho}=\sum_s\fal{\lambda}_s\fal{\rho}_s^{II}}

\newcommand{\rosep}{\rho=\sum_i\lambda_i\rho_i^I\oti\rho_i^{II}}
\newcommand{\rosepn}{\rho=\sum_{i=1}^N\lambda_i\rho_i^I\oti\rho_i^{II}}
\newcommand{\rosepfal}{\rho=\sum_{j=1}^{\fal{N}}\fal{\lambda}_j\fal{\rho}_j^I\oti\fal{\rho}_j^{II}}

\newcommand{\rosepf}{\rho=\sum_j\fal{\lambda}_j\fal{\rho}_j^I\oti\fal{\rho}_j^{II}}
\newcommand{\roseph}{\rho=\sum_j\widehat{\lambda}_j\widehat{\rho}_j^I\oti\widehat{\rho}_j^{II}}
\newcommand{\rosepfo}{\sum_j\fal{\lambda}_j\fal{\rho}_j^I\oti\fal{\rho}_j^{II}}
\newcommand{\korelsep}{\korel=\sum_i\lambda_i\bla F\bra_{\rho_i^I}\bla G\bra_{\rho_i^{II}}}
\newcommand{\korels}{\sum_i\lambda_i\bla F\bra_{\rho_i^I}\bla G\bra_{\rho_i^{II}}}
\newcommand{\korelsepf}{\korel=\sum_j\fal{\lambda}_j\bla F\bra_{\fal{\rho}_j^I}\bla G\bra_{\fal{\rho}_j^{II}}}
\newcommand{\korelseph}{\korel=\sum_j\widehat{\lambda}_j\bla F\bra_{\widehat{\rho}_j^I}\bla G\bra_{\widehat{\rho}_j^{II}}}

\newcommand{\Snd}{S^{nd}}

\newcommand{\cA}{{\cal A}}
\newcommand{\cB}{{\cal B}}
\newcommand{\cC}{{\cal C}}

\newcommand{\cH}{{\cal H}}

\newcommand{\cL}{{\cal L}}

\begin{document}

\maketitle

\begin{abstract}
A model of evolution of bipartite quantum state entanglement is studied. It involves the recently introduced quantum block
spin-flip dynamics on a lattice. We find that for initially separable states the considered evolution leads, in general, to
entangled states. We also present a complete characterization of two-point correlation functions for that type of dynamics
to confirm enhancement of quantum correlation for the considered system.
\end{abstract}

\section{Introduction}

Entanglement\footnote{We refer the reader not familiar with the notion of entanglement to Appendix A, where some
basic facts and definitions concerning separable and entangled states has been collected} first noted by Schr\"{o}dinger
and von Neumann as the characterictic trait of quantum mechanics \cite{N1},\cite{N2} and by Einstein, Podolsky, Rosen
\cite{N3} as a quantum superposition of two distinct states of physical systems with the great impact on our
understanding of the notion of reality in the atomic scale, is undoubtedly an essential feature of quantum mechanics.
That concept can be relatively easy treated for the hamiltonian type evolution. Namely, writing the full Hamiltonian of a
composite system, one should specify the interaction part explicitly. We want to emphasize that it is exactly that part
of hamiltonian, which is responsible for the evolution of entanglement in the following sense: it is impossible that all
factorizable states remain factorizable during the interaction unless the interaction part of the hamiltonian is trivial.

The question of the evolution of entanglement for quantum stochastic semigroups is much harder task. To explain that point
in detail, we recall that in recent papers, B. Zegarlinski and one of us, proposed a general scheme for a quantization of
stochastic dynamics which describe interacting classical particles \cite{R7}-\cite{R10}. In that scheme, guided by
classical theory (cf. \cite{R1}), a general recipe for quantum stochastic dynamics of jump type and diffusive type was
given (see also \cite{R4},\cite{R5},\cite{R6},\cite{N6}). In particular, to define the infinitesimal generators of Markov
semigroups, the theory of generalized conditional expectations (in the Accardi-Cechnini sense) in the framework of
quantum $L_p$-spaces was used. In that way, a general scheme to produce, describe and analyze dynamical systems with
evolution originating from quantization of stochastic processes and such that their equilibrium states are prescribed
(quantum) Gibbs states, was established.

Having such a general plan of quantization of stochastic dynamics one may pose a natural question about its
nontriviality. Under this notion we understand, first of all, that infinitesimal generator of such dynamics is not a
function of the hamiltonian which can be extracted from the given Gibbs state. This requirement arises in a natural way
from the methodology of construction of quantum stochastic dynamics sketched in the preceding paragraph. In fact, it has
been shown \cite{R11} that generators defined within the presented $L_p$-space setting satisfy the above specification. On
the other hand, the genuine quantum map should produce quantum correlations besides being well defined in non-commutative
structures. The important point to note here is the idea of quantum correlations. Recently, the concept of coefficient of
quantum correlations was introduced \cite{N5} and it was shown that such coefficient is not equal to zero only for
non-separable states. In other words the entanglement is closely related to non-classical correlations between two
subsystems of a composite system and such correlations arise from nontrivial interactions between the subsystems.
Therefore, to have nontrivial quantum dynamics one should show that the dynamical maps under considerations are able to
increase entanglement.

The main difficulty in showing an increment of quantum correlations stems from the fact that the explicit form of
interactions responsible for the transition rates used in the definition of quantum Markov generator is not known.
Consequently, the present case is much more difficult from the hamiltonian one. To start a general analysis of
evolution of entanglement for quantum stochastic dynamics, we begin with the block-spin flip type dynamics, which is
the main object of interest in this paper. Let us add that this dynamics can serve as a paradigm for a quantization of
Glauber dynamics. Thus, we will be concerned with evolution of entanglement in concrete finite dimensional model of
quantum block-spin flip dynamics.

In order to carry out the analysis of entanglement we use two different approaches. Having fixed notations and given
preliminaries (Section II), in Section III we consider an explicit example of low-dimensional jump-type system, showing
that the block spin flip dynamics leads to the entanglement of the initial separable quantum state. In Section IV we
study some properly chosen correlation functions and show that an enhancement of quantum correlation is {\em typical}
for the considered dynamics (see Proposition \ref{tw6}). Finally, in Section V we will comment the obtained results.

\section{Quantum block spin-flip dynamics}

In the general approach to quantum jump-type dynamics of quantum systems on a lattice it is convenient to consider
firstly the finite volume case - a system associated with a finite region $\Lambda$, and then to perform the
thermodynamic limit with $\Lambda$ going to ${\bf Z}^d$, where ${\bf Z}^d$ denotes $d$-dimensional lattice (for
mathematical details of algebraic description of quantum statistical mechanics see \cite{R2}). However, as we have
mentioned, the main scope of the paper is the analysis of evolution of entanglement and increment of entanglement is
taken as a signature of nontrivial interactions between two subsystems. This phenomenon should be present both for finite
and infinite subsystems. To simplify our exposition as much as possible we restrict to the essential ingredients of the
finite volume case (for a general description see \cite{R7},\cite{R9},\cite{R10}). Thus, we shall consider a composite
system $I+II$ associated with a region $\Lambda=\Lambda_I\cup\Lambda_{II}$, where $\Lambda_i$, $i=I,II$ are disjoint
finite subregions of the lattice ${\bf Z}^d$. To have a concrete dynamical system we will describe the construction of
the block spin-flip dynamics related to the region $\Lambda$. To this end, we associate with $\Lambda_I$ ($\Lambda_{II}$)
the finite dimensional Hilbert space $\cH_{\Lambda_I}\equiv\hj$ ($\cH_{\Lambda_{II}}\equiv\hd$) as the space of its pure
states, the set of density matrices $S_1$ ($S_2$) as the space of all mixed states, and the set of all bounded linear
operators $\cB(\hj)$ ($\cB(\hd)$) as the algebras of observables. Thus, the composite system $\Lambda$ is described by
$\hjd$, $S_1\oti S_2$ and $\cB(\hj)\oti\cB(\hd)\cong\cB(\hjd)$, respectively.

The reference state, playing the crucial role for classical and quantum case, is given here by the Gibbs state, i.e.
with each region $\Lambda$ ($\Lambda_I,\Lambda_{II}$, respectively) we associate the Hamiltonian $H_{\Lambda}$
($H_{\Lambda_I},H_{\Lambda_{II}}$, respectively) and subsequently the corresponding Gibbs state
$\omega_{\Lambda}(\cdot)=Tr\rho_{\Lambda}(\cdot)$ with $\displaystyle{\rho_{\Lambda}=\frac{e^{-\beta H_{\Lambda}}}{Tr
e^{-\beta H_{\Lambda}}}\equiv\rho}$, etc.

Guided by the classical theory (cf. \cite{R1}), where conditional expectations serve for the construction of jump type
stochastic processes, we use their non-commutative generalizations (in the Accardi-Cechini sense) to define the
infinitesimal generator $\cL_{\Lambda,\Lambda_I}\equiv \cL$ of the corresponding quantum spin-flip semigroup where the
"block spin flip" is carried out on $\Lambda_I$ ($\subseteq\Lambda$). Therefore let us introduce a map
$E_{\Lambda,\Lambda_I}(\equiv E):\cB(\hjd)\to\cB(\hjd)$ defined as follows:
\beqar
E(A)=Tr_1(\gamma^*A\gamma) \label{Edef}
\eeqar
where
\beqar
\gamma=\rho\prw\big(Tr_1\rho\big)\mprw \nonumber
\eeqar
with $Tr_1$ denoting the partial trace (over the Hilbert space $\hj$). Let us remark, that for infinite region case,
$\gamma$ is defined as analytic extension of non-commutative Radon-Nikodym cocycles (cf. \cite{R9}). Using the above
defined generalized conditional expectation $E$ we can introduce the following operator $\cL$ defined on $\cB(\hjd)$ by
\beqar
\cL(A)=E(A)-A \nonumber
\eeqar
for $A\in\cB(\hjd)$. {\em The important point to note here is the form of the infinitesimal Markov generator $\cL$: there
is no explicit term describing the interactions between regions $\Lambda_I$ and $\Lambda_{II}$} and this is the origin of
difficulties in the presented analysis. Given a state $\omega_{\rho}$, defined by a density matrix $\rho$,
$\omega_{\rho}=Tr(\rho(\cdot))$, one can define on $\cB(\hjd)$ the following scalar product
\beqar
\bla\bla A,B\bra\bra_{\omega_{\rho}}\equiv Tr\big(\rho\prw A^*\rho\prw B\big) \nonumber
\eeqar
Then, one can verify that $\big(\cB(\hjd),\la\la \cdot,\cdot\ra\ra_{\omega_{\rho}}\big)$ is the non-commutative Hilbert
space (which can be called the quantum Liouville space). It will be denoted by $L_2(\cB(\hjd),\rho)$. Moreover, one can
show that $\cL$ is a well defined bounded Markov generator such that
\beqar
\bla\bla \cL(A),B\bra\bra_{\omega_{\rho}}=\bla\bla A,\cL(B)\bra\bra_{\omega_{\rho}} \nonumber
\eeqar
It easily follows that the following semigroup $T_t^{\Lambda}(\equiv T_t)=\exp(t\cL)$ {\em is a well defined Markov
uniformly continuous semigroup} such that it is self-adjoint on the non-commutative Hilbert space, the state
$\omega_{\rho}$ is invariant (with respect to $T_t$) and $T_t$ can be represented as the sum of the following
convergent series:
\beqar
{\mathcal{I}}+t\cL+\frac{t^2}{2!}\cL^2+\cdots \nonumber
\eeqar
with ${\mathcal{I}}$ being the identity operator.

\section{Evolution of entanglement}

In the next two sections we will look more closely at the time evolution of quantum correlations. Here, we will present
the simplest nontrivial example clearly showing that quantum dynamics can produce this type of correlations. We will
analyze a $2\times 2$ system with block spin-flip dynamics. Thus, $\hj=\Cn^2=\hd$ and the Hilbert space of the
composite system is given by $\Cn^2\oti\Cn^2\approx\Cn^4$. Let $\{\xi_1,\xi_2\}$ be an orthonormal basis in $\Cn^2$. We
define:
\beqar
& & x_1=\frac{1}{\sqrt{2}}\big(\xjj+\xdd\big) \nonu \\
& & x_2=\xjd \nonu \\
& & x_3=\xdj \label{bazadef} \\
& & x_4=\frac{1}{\sqrt{2}}\big(\xjj-\xdd\big) \nonu
\eeqar
One can easily check that $\{x_i\}_{i=1}^4$ forms the orthonormal basis in$\Cn^4$. Let us define a faithful density matrix
$\rho$ on$\Cn^4$ which is given by the formula:
\beq
\rho:=\sum_{i=1}^4\lambda_i\big|x_i\bra\bla x_i\big| \qquad \lambda_i>0, \quad\sum_{i=1}^4\lambda_i=1 \label{Rodef}
\eeq
The just defined $\rho$ will play the role of the reference state (cf. Introduction). We will need the following the
well known fact which can be verified by straightforward calculation.
\begin{stw}
Let $\hj,\hd$ be Hilbert spaces. For every $x,v\in\hj$ and $y,z\in\hd$ we have: $Tr_1\big(\big|x\oti y\bra\bla v\oti z\big|
\big)=(x,v)\big|y\bra\bla z\big|$, where $Tr_1$ is the partial trace over $\hj$ in $\hjd$ and $(\cdot,\cdot)$ denotes the
scalar product (in $\hj$). \label{P1}
\end{stw}
In the sequel we shall identify $Tr_1(\cdot)$ with its embedding into $\hjd$, defined as $\iden\otimes Tr_1(\cdot)$. 
Now, let us consider the quantum spin-flip type dynamics $T_t$ for our model. Its infinitesimal generator is defined as
(cf. Section II):
\beqar
\cL(f)=E(f)-f \nonu
\eeqar
where $f\in\cB(\hjd)$, while $E$ was defined in the previous section.

The dynamical semigroup $T_t$ has the additional property: as ${\mathcal{A}}\subset L_2(\mathcal{A},\rho)$, we have
$T_t\mathcal{A} \subset\mathcal{A}$ (Feller property), where $\cA=\cB(\hjd)$, $L_2(\cA,\rho)$ is the non-commutative
Hilbert space.

It follows that the Feller property allows to study the following duality problem: We may consider the time evolution
$T_t$ as the family of maps $T_t:\mathcal{A}\rightarrow\mathcal{A}$, then we can apply the standard equivalence between
Schr\"odinger and Heisenberg picture to determine the evolution $T_t^d$ of a state $\sigma$. To this end we define:
\beqar
Tr\big(T_t^d(\sigma)f\big):=Tr\big(\sigma T_t(f)\big) \nonu
\eeqar
for any state $\sigma$ and any observable $f$. Therefore, we are able to describe explicitly the time evolution of states
for that type of dynamics, which is given by the following mapping:
\beq
\sigma\;\rightarrow\; T_t^d(\sigma)=\sigma+t\bigg(E^d(\sigma)-\sigma\bigg)+\cdots \label{kropki1}
\eeq
where the dual $E^d$ of the infinitesimal generator $E$ is defined by the equality:
\beqar
Tr\big(E^d(\sigma)f\big)=Tr\big(\sigma E(f)\big) \nonu
\eeqar
and the series in the right hand side of (\ref{kropki1}) is convergent. Using (\ref{Edef}) we can write:
\beqar
Tr\big(E^d(\sigma)f\big)=Tr\big(\sigma Tr_1(\gamma^*f\gamma)\big)=Tr\big(Tr_1(\sigma)\gamma^*f\gamma\big)= Tr\big(\gamma
Tr_1(\sigma)\gamma^*f\big) \nonu
\eeqar
Thus, we get:
\beq
E^d(\sigma)=\gamma Tr_1(\sigma)\gamma^*=\rho\prw\big(Tr_1\rho\big)\mprw Tr_1\sigma\big(Tr_1\rho\big)\mprw\rho\prw \label{Ed}
\eeq
Let us put $\sigma=\sigma^I\oti\sigma^{II}$, and recall that $\rho$ was defined by (\ref{Rodef}). Obviously
\beq
Tr_1\sigma=\iden_I\oti\sigma^{II} \label{trsigma}
\eeq
Using Proposition \ref{P1} we can easily calculate $Tr_1(\rho)=\sum_{i=1}^4\lambda_i Tr_1|x_i\ra\la x_i|$. We have
\beqar
Tr_1|x_1\ra\la x_1|=\frac{1}{2}Tr_1|\xjj+\xdd\ra\la\xjj+\xdd|= \nonu
\eeqar
\beqar
\frac{1}{2}\big[Tr_1|\xjj\ra\la\xjj|+ Tr_1|\xjj\ra\la\xdd|+ \nonu
\eeqar
\beqar
+Tr_1|\xdd\ra\la\xjj|+Tr_1|\xdd\ra\la\xdd|\big]= \nonu
\eeqar
\beqar
\frac{1}{2}\big[\iden_I\oti\big(|\xi_1\ra\la\xi_1|+ \xi_2\ra\la\xi_2|\big)\big]=\frac{1}{2}\big[\iden_{II}\big] \nonu
\eeqar
For simplicity, we will denote $\iden_I$ and $\iden_{II}$ briefly by $\iden$ when no confusion can arise. Analogously:
\beqar
Tr_1|x_2\ra\la x_2|=Tr_1|\xjd\ra\la\xjd|=\iden\oti\rzdd \nonu
\eeqar
\beqar
Tr_1|x_3\ra\la x_3|=Tr_1|\xdj\ra\la\xdj|=\iden\oti\rzjj \nonu
\eeqar
\beqar
Tr_1|x_4\ra\la x_4|=\frac{1}{2}Tr_1|\xjj-\xdd\ra\la\xjj-\xdd|=\frac{1}{2}\big[\iden\oti\iden\big] \nonu
\eeqar
Eventually, we obtain:
\beqar
Tr_1\rho=\frac{\lambda_1}{2}\iden+\lambda_2\iden\oti\rzdd+\lambda_3\iden\oti\rzjj+\frac{\lambda_4}{2}\iden= \nonu
\eeqar
\beqar
\bigg(\frac{\lambda_1+\lambda_4}{2}+\lambda_3\bigg)\iden\oti\rzjj+\bigg(\frac{\lambda_1+\lambda_4}{2}+\lambda_2\bigg)
\iden\oti\rzdd \label{trro}
\eeqar
We introduce the following notation:
\beqar
\chi_1=\bigg(\frac{\lambda_1+\lambda_4}{2}+\lambda_3\bigg)^{-1}\qquad
\chi_2=\bigg(\frac{\lambda_1+\lambda_4}{2}+\lambda_2\bigg)^{-1} \nonu
\eeqar
Inserting (\ref{Rodef}), (\ref{trsigma}), and (\ref{trro}) into (\ref{Ed}) we get:
\beqar
E^d(\sigma)=\bigg(\sum_{i=1}^4\lambda_i\prw|x_i\ra\la
x_i|\bigg)\cdot\bigg(\iden\oti\big[\chi_1\prw\rzjj+\chi_2\prw\rzdd\big]\bigg) \cdot \nonu
\eeqar
\beqar
\cdot\big(\iden\oti\sigma^{II}\big)\cdot\bigg(\iden\oti\big[\chi_1\prw\rzjj+\chi_2\prw\rzdd\big]\bigg)
\cdot\bigg(\sum_{i=1}^4\lambda_i\prw|x_i\ra\la x_i|\bigg) \nonu
\eeqar
Now, suppose that $\sigma^{II}=a\rzjj+b\rzdd$ with $a\geq 0, b\geq 0, a+b=1$. Then,
\beqar
E^d(\sigma)=\bigg(\sum_{i=1}^4\lambda_i\prw|x_i\ra\la x_i|\bigg)\cdot\big(\iden\oti a\chi_1\rzjj+\iden\oti
b\chi_2\rzdd\big)\cdot\bigg(\sum_{i=1}^4\lambda_i\prw|x_i\ra\la x_i|\bigg) \nonu
\eeqar
Using (\ref{bazadef}) and performing some lengthy calculation one can obtain:
\beqar
E^d\big(\sigma^I\oti(a\rzjj+b\rzdd)\big)= \bigg[a\chi_1\big(\lambda_1\prw+\lambda_4\prw\big)^2+b\chi_2
\big(\lambda_1\prw-\lambda_4\prw\big)^2\bigg]\rzjj\oti\rzjj+ \nonu
\eeqar
\beqar
+\big(a\chi_1+b\chi_2\big)\big(\lambda_1-\lambda_4\big) \rzjd\oti\rzjd+
\big(a\chi_1+b\chi_2\big)\big(\lambda_1-\lambda_4\big) \rzdj\oti\rzdj+ \nonu
\eeqar
\beqar
+\bigg[a\chi_1\big(\lambda_1\prw-\lambda_4\prw\big)^2+b\chi_2 \big(\lambda_1\prw+ \lambda_4\prw\big)^2\bigg]\rzdd\oti\rzdd+
\nonu
\eeqar
\beqar
+a\chi_1\lambda_3\rzdd\oti\rzjj+b\chi_2\lambda_2\rzjj\oti\rzdd \nonu
\eeqar
Now, we assume that $\lambda_2=\lambda_3$ which implies $\chi_1=\chi_2=\chi$. Then,
\beqar
E^d\big(\sigma^I\oti(a\rzjj+b\rzdd)\big)= \chi\bigg[\big[a\big(\lambda_1\prw+\lambda_4\prw\big)^2+b
\big(\lambda_1\prw-\lambda_4\prw\big)^2\big]\rzjj\oti\rzjj+ \nonu
\eeqar
\beqar
+\big(\lambda_1-\lambda_4\big) \rzjd\oti\rzjd+ \big(\lambda_1-\lambda_4\big) \rzdj\oti\rzdj+ \nonu
\eeqar
\beqar
+\big[a\big(\lambda_1\prw-\lambda_4\prw\big)^2+b \big(\lambda_1\prw+ \lambda_4\prw\big)^2\big]\rzdd\oti\rzdd+ \nonu
\eeqar
\beqar
a\lambda_2\rzdd\oti\rzjj+b\lambda_2\rzjj\oti\rzdd\bigg] \nonu
\eeqar
Performing some easy but tedious calculations we arrive at the following decomposition:
\beqar
E^d(\sigma)=\sum_{i=1}^4\fal{\lambda}_i\big|y_i\bra\bla y_i\big| \nonu
\eeqar
with
\beqar
\fal{\lambda}_1=\chi\lambda_+\qquad \fal{\lambda}_2=\chi b\lambda_2\qquad \fal{\lambda}_3=\chi a\lambda_2\qquad
\fal{\lambda}_4=\chi\lambda_- \nonu
\eeqar
where
\beqar
\lambda_{\pm}=\frac{A+C\pm X}{2}\qquad X=\sqrt{(A-C)^2+4B^2} \nonu
\eeqar
and
\beqar
A:=a\big(\lambda_1\prw+\lambda_4\prw\big)^2+b \big(\lambda_1\prw-\lambda_4\prw\big)^2\qquad B:=\lambda_1-\lambda_4 \qquad
C:=a\big(\lambda_1\prw-\lambda_4\prw\big)^2+b \big(\lambda_1\prw+\lambda_4\prw\big)^2\qquad \nonu
\eeqar
while $\{y_i\}_{i=1}^4$ is the orthonormal basis defined as below:
\beqar
& & y_1=\eta_+\xjj+\kappa_+\xdd \nonu \\
& & y_2=\xjd \nonu \\
& & y_3=\xdj \label{bazadsssef} \nonu \\
& & y_4=\eta_-\xjj+\kappa_-\xdd \nonu
\eeqar
where
\beqar
\eta_{\pm}:=\frac{\sqrt{2}B}{\sqrt{X^2\mp(A-C)X}}\qquad \kappa_{\pm}:=\frac{-(A-C)\pm X}{\sqrt{2}\sqrt{X^2\mp(A-C)X}} \nonu
\eeqar
The above decomposition of $E^d(\sigma)$ is well defined for $\lambda_1>\lambda_4$. In particular, we have
$\lambda_{\pm}>0$, $\eta_{\pm},\kappa{\pm}\neq 0$. \\
Now, we are in position to examine the separability of the state $E^d\big(\sigma^I\oti(a\rzjj+b\rzdd)\big)$. We will use
the simple argument presented in \cite{R18}. Define:
\beqar
E^d_0(\sigma)=\fal{\lambda}_1|y_1\ra\la y_1|+\fal{\lambda}_4|y_4\ra\la y_4| \nonumber
\eeqar
We observe
\beqar
E^d_0(\sigma)y_2=0=E^d_0(\sigma)y_3 \nonumber
\eeqar
and
\beqar
\big(E^d_0(\sigma)\xjj,\xdd\big)=\fal{\lambda}_1\eta_+\kappa_++\fal{\lambda}_4\eta_-\kappa_- \equiv A\nonumber
\eeqar
Let us put
\beqar
S=\sum_j A_j\oti B_j \nonumber
\eeqar
where $A_j$ ($B_j$) are positive operators in $\cB(\hj)$ ($\cB(\hd)$). Then,
\beqar
||E^d_0(\sigma)-S||\geq A-\sum_j||A_j^{\prw}\xi_1||\,||B_j^{\prw}\xi_2||\,||A_j^{\prw}\xi_2||\,||B_j^{\prw}\xi_1|| \nonumber
\eeqar
Hence,
\beqar
||E^d_0(\sigma)-S||\geq \frac{A}{2} >0 \nonumber
\eeqar
Consequently, $E^d_0(\sigma)$ is not an element of the closure of separable states. To show that
$(1-t)\sigma+tE^d(\sigma)$ is an entangled state we start with recalling some well known facts from the theory of
partially ordered topological spaces. Let us consider $\cA_{sa}=\{a\in\cA: a=a^*\}$, where
$\cA=\cB(\hjd)=\cB(\hj)\otimes\cB(\hd)$ as a real finite dimensional Banach space ($\hj$ and $\hd$ were assumed to be
finite dimensional Hilbert spaces). The set $\{conv(\cB_+(\hj)\otimes\cB_+(\hd))\}^{closure}$ will be denoted by $V$.
Clearly, $V$ is a proper generating cone in $\cA$. Further, by general argument (or by direct proof) one can note that
$\textrm{int} V\neq\phi$ (where int stands for interior of the set). Define $V_1=V\cap K(0,1)$, where
$K(0,1)=\{x\in\cA:||x||\leq 1\}$. One can verify that $V_1$ is convex, compact subset with non-empty interior.
Therefore, it is homeomorphic to unit ball, while its boundary $\partial V_1$ is homeomorphic to unital sphere. Thus,
it follows that, in general, a convex combination of $\rho_1\in V_1$ and $\rho_2\notin V_1$ is not in $V_1$. By
applying the above facts to the case $\rho_1=(1-t)\sigma$, $\rho_2=tE^d(\sigma)$, and taking into account the
equivalence of the trace and operator norms in finite dimensional case, one can draw the conclusion that within the
perturbation calculus in the first order a separable state $\sigma$ evolves to the entangled state
$(1-t)\sigma+tE^d(\sigma)$.

\section{Factorization of two-points correlation functions}

In this section we will proceed with analysis of factorization of certain correlation functions. To clarify the relation
of that topic to evolution of entanglement we start with establishing the relation between factorization and existence of
non-quantum correlation between two subsystems $\cA_1,\cA_2$. Having this as well as the relation between quantum
correlations and entanglement, we give a detailed analysis of factorization of two-point correlation functions.

Let us begin with a classical case, i.e. let $\cA_i$, $i=1,2$, be two abelian $C^*$-algebras with identities. Consider
$\cA_{cl}=\cA_1\otimes\cA_2$ and let $\omega:\cA_{cl}\to\Cn$ be a state. The abelianess of $\cA_i$, $i=1,2$, implies
\beqar
\cA_{cl}\cong \cC(\Omega_1)\otimes\cC(\Omega_2)=\cC(\Omega_1\times\Omega_2), \nonumber
\eeqar
where $\Omega_i$, $i=1,2$ are compact Hausdorff topological spaces. By Markov-Riesz theorem there exists a probability
measure $\mu$ on $\Omega_1\times\Omega_2$ such that
\beqar
\omega(a_1\otimes a_2)=\mu(\hat{a}_{12}), \nonumber
\eeqar
with $\hat{a}_{12}=i(a_1\otimes a_2)$ and $i$ being an isomorphism between $\cC(\Omega_1)\otimes\cC(\Omega_2)$ and
$\cC(\Omega_1\times\Omega_2)$. Consider the truncated correlation (or equivalently the second Ursell function)
\beqar
C_{\omega,a_1,a_2}\equiv\omega(a_1\otimes\iden\cdot\iden\otimes a_2)-\omega(a_1\otimes\iden)\omega(\iden\otimes a_2)
\nonumber
\eeqar
for $a_i\in\cA_i$, $i=1,2$. Due to the fact that each (classical) measure can be $^*$-weakly approximated by finitely
supported probability measures one has
\beqar
C_{\omega,a_1,a_2}=\lim_n\left[\mu_n(\hat{a}_1\cdot\hat{a}_2)-\mu_n(\hat{a}_1)\mu_n(\hat{a}_2)\right], \nonumber
\eeqar
where for each $n$, $\mu_n=\sum_{i=1}^{N_n}\lambda_i^{(n)}\delta_{\omega_i^{(n)}}$, and $\delta_{\omega_i^{(n)}}$ is a
Dirac (point) measure on $\Omega_1\times\Omega_2$. Therefore
$\delta_{\omega_i^{(n)}}=\delta_{\omega_i^{(n,1)}}\times\delta_{\omega_i^{(n,2)}}$, where we have used the following
notation: $\Omega_1\times\Omega_2\ni\omega_i^n=(\omega_i^{(n,1)},\omega_i^{(n,2)})$. Consequently
\beqar
C_{\omega,a_1,a_2}=\lim_n\left[\sum_{i=1}^{N_n}\lambda_i^{(n)}\delta_{\omega_i^{(n,1)}}(\hat{a}_1)
\delta_{\omega_i^{(n,2)}}(\hat{a}_2) -\left(\sum_{i=1}^{N_n}\lambda_i^{(n)}\delta_{\omega_i^{(n,1)}}(\hat{a}_1)\right)
\left(\sum_{i=1}^{N_n}\lambda_i^{(n)}\delta_{\omega_i^{(n,2)}}(\hat{a}_2)\right)\right]. \nonumber
\eeqar
Thus, each classical truncated correlation $C_{\omega,a_1,a_2}$ can be approximated by a difference of "separable
states" and product of one-point functions. The important point to note here is that from the very beginning it is
necessary to determine subalgebras (here $\cA_1,\cA_2$) - then we can define correlations with respect to this fixed
partition. We recall that we are studying block spin-flip dynamics, where the spin-flip is carried out over the region
$\Lambda_I$. Thus, the partition was fixed. Moreover, by definition, a separable state has the factorization property!

Now, let $\cA_i$ be arbitrary (non-commutative) $C^*$-algebras with identities. Let $\omega$ be a separable state on
$\cA=\cA_1\otimes\cA_2$, where in our case $\cA=\cB(\hj)\otimes\cB(\hd)$, $\cA_i=\cB(\cH_i)$, $i=1,2$. Then, guided by
the above general observation concerning classical systems, we say that a separable state
$\omega=\sum_i\lambda_i\rho_i^I\otimes\rho_i^{II}$ encodes classical correlations with respect to the partition of
$\cA=<\cA_1\otimes\iden,\iden\otimes\cA_2>$, a $C^*$-algebra generated by $\cA_1\otimes\iden$ and $\iden\otimes\cA_2$.

Again, let $\omega$ be a separable state with respect to the partition $\cA=<\cA_1\otimes\iden,\iden\otimes\cA_2>$ and
$T_t:\cA\to\cA$ be a spin-flip type dynamics. We wish to consider the truncated correlation (now for our quantum
dynamical system)
\beqar
C_{\omega,g,f}^{q,T}=\omega(gT_t(f))-\omega(g)\omega(T_t(f)) \nonumber
\eeqar
with $g\in id_{\hj}\otimes\cA_2$, $f\in\cA_1\otimes id_{\hd}$. Throughout the rest of the paper we shall assume that
observables $f$ and $g$ are of the form:
\beqar
\fr,\qquad\gr \nonu
\eeqar
with $F\in{\mathcal{B}}(\hj)$ and $G\in{\mathcal{B}}(\hd)$, i.e. $f$ ($g$) is an element of the subsystem $I$ ($II$
respectively). We observe:
\beqar
C_{\omega,g,f}^{q,t}=\omega(gT_t(f))-\omega(g)\omega(T_t(f))=\omega(gf)+t\omega(g\cL(f))+\frac{t^2}{2!}\omega(g\cL(\cL(f)))
+\ldots -\omega(g)\omega(f) \nonumber
\eeqar
\beqar
=C_{\omega,g,f}^{q}+tC_{\omega,g,f}^{L,1}+\frac{t^2}{2!}C_{\omega,g,f}^{L,2}+\frac{t^3}{3!}C_{\omega,g,f}^{L,2}+\ldots
\nonumber
\eeqar
where
\beqar
C_{\omega,g,f}^{q}=\omega(gf)-\omega(g)\omega(f) \nonumber
\eeqar
\beqar
C_{\omega,g,f}^{L,1}=\omega(g\cL(f))- \omega(g)\omega(\cL(f)), \quad C_{\omega,g,f}^{L,2}=\omega(g\cL(\cL(f)))-
\omega(g)\omega(\cL(\cL(f))), \quad \ldots \nonumber
\eeqar
We observe:

i) $C_{\omega,g,f}^{q}$ measures the "classical" correlations between $g$ and $f$ with respect to the partition $\cA_1$,
$\cA_2$, for time equal to $0$.

ii) We can not say the same for $C_{\omega,g,f}^{L,1}$, $C_{\omega,g,f}^{L,2}$, etc., since $\omega$ is separable with
respect to the partition $\cA_1\otimes\iden$, $\iden\otimes\cA_2$, while $g\in\cA_2$ but $\cL(f)\in\cA$! We would have
"classical" correlations if $\omega$ was of the form $\omega=\sum_i\lambda_i\omega_i^I\otimes\omega_i^{II}$, where
$\omega_i^I$ is a state on a $C^*$-subalgebra containing $\cL(f)$ while $\omega_i^{II}$ is a state on $\cA_2$, and
similarly for higher order terms. In other words, the evolution (its infinitesimal generator) leads to a deviation from
the original partition of the composite system and the new partition does not fit to our definition of the spin-flip
operation over the region $I$. In particular, the given separable decomposition of $\omega$ is not adapted to be a
measure of classical correlations between $g$ and $\cL(f)$.

Therefore, nonzero value of $\omega(G\cL(F))\equiv C_{\omega,g,f}^{L,1}$ can be taken as an indicator for the increment
of non-classical correlations (with respect to the partition $<\cA_1\otimes\iden,\iden\otimes\cA_2>$).

To elaborate that point a little bit further let the state $\omega$ be given by the density matrix $\rho$ of the form
\beqar
\rho=\sum_i\lambda_i\rho^I_i\otimes\rho^{II}_i. \label{postac_rho}
\eeqar
with $\rho_i^I$ and $\rho_i^{II}$ being the states of the subsystem $I$ and $II$, respectively. Consider
$\omega(GE(F))=\sum_i\lambda_iTr\Big(\rho^I_i\otimes\rho^{II}_i\cdot G\cdot E(F)\Big)$ with $G\in\cA_2$, $F\in\cA_1$
(with a natural embedding of $\cA_1$ and $\cA_2$ into $\cA$). Assume $G=HH^*$ with $[H,\rho_i^{II}]=0$. Then,
\beqar
\omega(GE(F))=\sum_i\lambda_iTr\rho^I_i\otimes H\rho^{II}_iH^* E(F)=
\sum_i\lambda_i\rho_i^{II}(HH^*)Tr\rho_i^I\otimes\rho_i^{II,H}E(F)= \nonumber
\eeqar
\beqar
=Tr E^d\Big(\sum_i\lambda_i\rho_i^{II}(HH^*)\rho_i^I\otimes\rho_i^{II,H}\Big)\cdot F \nonumber
\eeqar
with $\rho_i^{II,H}=\rho_i^{II}/\rho_i(H^*H)$. We know, by Section III, that
$E^d\Big(\sum_i\lambda_i\rho_i^{II}(HH^*)\rho_i^I\otimes\rho_i^{II,H}\Big)$ is not, in general, a separable state. In
other words, any reasonable factorization does not hold. In particular,
\beqar
E^d\Big(\sum_i\lambda_i\rho_i^{II}(HH^*)\rho_i^I\otimes\rho_i^{II,H}\Big)(F\otimes id_{\hd})
\neq\sum_i\lambda_i\rho_i^{II}(HH^*)\rho_i^I(F) \nonumber
\eeqar
We can summarize the above consideration as follows:

{\em An entangled state encodes non-classical correlations. Even in the simplest case of $2\times 2$ system the spin-flip
dynamics can lead to entangled state (cf. Section III). Moreover, in general, for the entangled state, the second Ursell
function fails to have the factorization property. }

Before performing the promised analysis we want to make an additional observation, which justifies our assumption that
it is enough to consider only observables from subalgebras while considering correlation function $\omega(GE(F))$.
\begin{uw}
We note that the equality
\beqar
0=\omega(G\cL(F))=\omega(GE(F))-\omega(GF) \label{pom1}
\eeqar
for all $G\in\cA$ and $F\in\cA_2$ is a very strong condition, since it implies $E=id$, so the spin-flip dynamics would be
a trivial one. Therefore we restrict ourselves to study much weaker condition
\beqar
\omega(GE(F))\neq\omega(GF) \nonumber
\eeqar
for $G\in\cA_1$, $F\in\cA_2$ and a separable state $\omega$ on $\cA_1\otimes\cA_2$. \label{uwaga_korel1}
\end{uw}
Consequently, we shall analyze conditions under which the function $<G,F>\to\omega(GE(F))$ does not factorize, i.e.
(\ref{pom1}) does not hold for $G\in\cA_1$, $F\in\cA_2$. Again, we assume that the state $\rho$ is of the form
(\ref{postac_rho}). From now on we make the assumption that the density matrix $\rho$ is an invertible one. This
assumption stems from the general strategy of constructing quantum maps. Namely, we associate with the quantum system
the quantum Hilbert space $L_2(\cA,\rho)$ with a given reference Gibbs state. Obviously, any Gibbs state has the
assumed property. We also note that if $\rho$ is an invertible density matrix, then $Tr_1(\rho)$ has this property too.
This observation will be used throughout this section.

As a result of longish (however not difficult)
calculations we get the following characterization of the considered correlation functions: \\
\emph{ Let $\bazajd$ be the orthonormal basis of $\hjd$, where $\bazaj$ is arbitrary orthonormal basis of $\hj$ while the
basis $\bazad$ in $\hd$ is such that $Tr_1(\rho)$ is diagonal. Then we have:
\beqar
\korel=\sum_{klji}\bigg(\sum_{pq}\rho\prw_{pkjq}\rho\prw_{lpqi}\sqrt{\frac{\fal{a}_j}{\fal{a}_i}}\bigg)
\bla\varphi_k\big|F\big|\varphi_l\bra\bla\phi_i\big|G\big|\phi_j\bra \label{E*1}
\eeqar
$\qquad$ $\qquad$ $\qquad$ where
\beqar
\fal{a}_j=\sum_r\rho_{rrjj} \nonu
\eeqar
$\qquad$ $\qquad$ $\qquad$ and the matrix elements in the basis $\bazajd$ are:
\beqar
\rho_{pqrs}:=\bla\varphi_p\oti\phi_r\big|\rho\big|\varphi_q\oti\phi_s\bra \qquad
\rho_{pqrs}\prw:=\bla\varphi_p\oti\phi_r\big|\rho\prw\big|\varphi_q\oti\phi_s\bra \label{E*3}
\eeqar
Moreover, $\fal{a}_j$ are the eigenvalues of $Tr_1(\rho)$. \label{fakt1}}

\subsection{General characterization of factorization}

In order to examine the just described correlation functions, we assume as before that $\rho$, $\rho^I$ and $\rho^{II}$
(with or without indexes) denote states on $\hjd$, $\hj$ and $\hd$, respectively.

It is evident from the definition of $C_{\omega,f,g}^{L,1}$ that the factorization of the correlation function $\korel$
is not dependent on a particular choice of the decomposition of $\rho$. This allows us to examine the factorization for
any decomposition of $\rho$ with conclusions valid for any other decomposition. As the next step of mathematical
framework for characterizing of correlation functions, we want to give the necessary and sufficient condition for the
factorization of such functions.
\begin{stw}
Let $\rosep$ and $\fr$, $\gr$, $dim\hj=n$, $dim\hd=m$. Let $\bazaj\si$ be arbitrary orthonormal basis of $\hj$ and
$\bazad\sj$ be orthonormal basis of $\hd$ such
that $\fal{\rho}\equiv Tr_1\rho$ is diagonal. Then the following conditions are equivalent: \\
(i) correlation function can be factorized, i.e.
\beqar
\korelsep \nonu
\eeqar
(ii) for every $\;k,l\in\{1,2,\ldots,n\}$ and $\;j,i\in\{1,2,\ldots,m\}$ the following equality holds:
\beqar
\rho_{lkji}=\sum_{p=1}^n\sum_{q=1}^m\rho\prw_{pkjq}\rho\prw_{lpqi} \sqrt{\frac{\sum_{r=1}^n\rho_{rrjj}}
{\sum_{v=1}^n\rho_{vvii}}} \label{tw2:ii}
\eeqar
where $\rho_{lkji}:=\bla\varphi_l\oti\phi_j\big|\rho\big|\varphi_k\oti\phi_i\bra$. \\
If the condition (ii) holds for some basis $\bazaj$ then it holds also for any other basis $\{\fal{\varphi}_i\}$ (basis
$\bazad$ unchanged!). Conversely, if (ii) does not hold for given basis $\bazaj$ then it does not hold for any other basis
$\{\fal{\varphi}_i\}$. \label{tw2}
\end{stw}
\dw Take the bases $\{\varphi_k\}$, $\bazad$ such as described in the proposition. Calculate $\korels$ in the basis
$\{\varphi_k\oti\phi_j\}$. Using (\ref{E*3}), we get:
\beqar
\korels=\sum_i\lambda_i\sum_l\bla\varphi_l\big|\rho_i^IF\big|\varphi_l\bra\sum_j\bla\phi_j\big|\rho_i^{II}G\big|\phi_j\bra
=\sum_{kljr}\rho_{lkjr}\bla\varphi_k\big|F\big|\varphi_l\bra\bla\phi_r\big|G\big|\phi_j\bra \nonu \label{postl}
\eeqar
From the general form of the considered correlation function (cf. (\ref{E*1})) we have:
\beqar
\korel=\sum_{kljr}\bigg(\sum_{pq}\rho\prw_{pkjq}\rho\prw_{lpqr}\sqrt{\frac{\fal{a}_j}{\fal{a}_r}}\bigg)
\bla\varphi_k\big|F\big|\varphi_l\bra\bla\phi_r\big|G\big|\phi_j\bra \quad \textrm{where} \quad \fal{a}_j=\sum_v\rho_{vvjj}
\label{postp} \nonu
\eeqar
Substituting $F=\big|\varphi_k\bra\bla\varphi_l\big|$ and $G=\big|\phi_j\bra\bla\phi_i\big|$ with $k,l,j,i$ arbitrary, it
is easy to verify that the right-hand sides of the last two expressions equal if and only if (\ref{tw2:ii}) holds. \finito

\subsection{Factorization and quasi-classicality} \label{sekcjaB}

It turns out that the sufficient conditions for factorization of correlation function can be connected to the
'quasi-classicality' of the considered state - the notion to be precised in the following definition. If $\rosep$ is given
decomposition of a separable density matrix, then $\rho$ can be considered as classical if $\{\rho^I_i\}$ and
$\{\rho^{II}_i\}$ are abelian families of density matrices. Below, we define weaker conditions for families of density
matrices, which are essential for the subsequent considerations.
\begin{deff}
Let $\cH$ be the Hilbert space, $dim\cH=n$. Let $\{\rho_i\}_{i=1}^N$ be a family of density matrices on $\cH$ and
$\{\lambda_i\}_{i=1}^N$ be (strictly) positive numbers $(\lambda_i>0)$, such that $\sum_{i=1}^N\lambda_i=1$. Define
$\rho_0:=\sum_i\lambda_i\rho_i$. Let $\{\phi_j\}_{j=1}^n$ - orthonormal basis consisting of eigenvectors of $\rho_0$ and
$\{c_j\}_{j=1}^n$ - corresponding eigenvalues. We will say that the family $\{\rho_i\}$ is \udl{K-quasi-abelian ($1\leq
K\leq n$)} if and only if the following condition holds:
there exists the partition $\{A_p\}_{p=1}^K$ of $\{1,2,\ldots,n\}$, $|A_p|\geq 1$ (where $|A|$ denotes the cardinality
of the set $A$), such that: \\
(i) for any $p\in\{1,\ldots,K\}$ and $r,s\in A_p$ we have $c_r=c_s$ and for any $r\in A_p,\;s\in A_q$ if $p\neq q$ then
$c_r\neq c_s$ \\
(ii) for any $k,l\in\{1,\ldots,n\}$ and $i\in\{1,\ldots,N\}$ if $c_k\neq c_l$ then $\bla\phi_k\big|\rho_i\big|\phi_l\bra=0$
\finito \label{def:quasikom}
\end{deff}
\begin{uw}
Definition \ref{def:quasikom} can be rephrased in a simple way as follows. The family $\{\rho_i\}$ is K-quasi-abelian if
and only if $\rho_i$ commutes with spectral projectors of $\rho_0$ (so with $\rho_0$ itself) for each $i=1,\ldots,N$.
However, we shall use properties (i) and (ii) in the sequel, so the original formulation is more convenient.
\end{uw}
It is an easy observation that if $|A_p|=1$ for some $p$, then the corresponding vector $\phi_p$ is the common
eigenvector for all the matrices $\rho_i$. This means that the number of sets $A_p$ in the partition $\{A_p\}$ with the
cardinality one equals the number of common eigenvectors for all the matrices $\rho_i$. In particular, if $\{\rho_i\}$ is
$K$-quasi-abelian
for $K=n$, then it is abelian in the traditional sense. \\
On the basis of the above definition we can formulate the sufficient conditions for factorization of the correlation
function using the above type of commutativity properties of the families $\{\rho^I_i\}$ and $\{\rho^{II}_i\}$.
\begin{stw}
Let $\rosep$ be a separable density matrix and $\fr$, $\gr$, $dim\hj=n$, $dim\hd=m$. Then the following implication holds:
\begin{displaymath} \left(
\begin{array}{l} \textrm{ There exists decomposition } \rosepfal \textrm{ such that one of the conditions is satisfied:} \\
(i)\quad \{\fal{\rho}_j^{II}\} \textrm{ is K-quasi-abelian $(K<n)$ and } \{\fal{\rho}_j^{I}\}
\textrm{ is abelian } \\
(ii)\quad \{\fal{\rho}_j^{II}\}\textrm{ is abelian }
\end{array}\right) \end{displaymath}
\begin{displaymath} \textrm{then}\quad \left( \begin{array}{l}
\korelsep \textrm{ (factorization of the correlation function)} \end{array} \right) \end{displaymath} \label{tw3}
\end{stw}
\dw We will prove the implication assuming (i). One can prove the statement under (ii) by similar reasoning. Let us
first prove that (\ref{tw2:ii}) holds. To this end let us take the bases $\bazaj$ and $\bazad$ such that
$\fal{\rho}_j^I$ are diagonal in the basis $\bazaj$ and $\sum_j\fal{\lambda}_j\fal{\rho}_j^{II}\equiv\trjr$ is diagonal
in the basis $\bazad$. Then, the matrix elements $\rho_{lkij}=0$ whenever $l\neq k$. The same is true for
$\rho\prw_{lkij}$. Note that also $\sum_{pq}\rho\prw_{pkjq}\rho\prw_{lpqi}\displaystyle{\sqrt{\fal{a}_j/\fal{a}_i}}=0$
for $l\neq k$ since for every $p,i,j$ and $q$ either $\rho\prw_{pkjq}=0$ or $\rho\prw_{lpqi}=0$. Hence, for $l\neq k$
(\ref{tw2:ii}) holds. Now, let $l=k$. Writing
$\sum_{pq}\rho\prw_{pkjq}\rho\prw_{lpqi}=\sum_p\big(\sum_q\rho\prw_{pkjq}\rho\prw_{lpqi}\big)$, we see that it can
differ from zero if and only if $p=k(=l)$. Taking into account that due to our specific choice of the bases, the
equality $\sum_q\rho\prw_{kkjq}\rho\prw_{kkqi}=\rho_{kkji}$ holds, we can write (\ref{tw2:ii}) for $l=k$ in the form
$\rho_{kkji}=\rho_{kkji}\sqrt{\fal{\rho}_{jj}/\fal{\rho}_{ii}}$ with
$\fal{\rho}_{jj}:=\bla\phi_j\big|\fal{\rho}\big|\phi_j\bra$, $\rozredf(\equiv\trjr)$. Note that because of the choice of
$\bazad$, elements $\fal{\rho}_{jj}$ are the eigenvalues of $\fal{\rho}$. By assumption, the family
$\{\fal{\rho}_j^{II}\}$ is K-quasi-abelian. This means that either $\fal{\rho}_{jj}=\fal{\rho}_{ii}$ ((\ref{tw2:ii}) is
then satisfied in an obvious way) or $\fal{\rho}_{jj}\neq\fal{\rho}_{ii}$. In the latter case we have
$\bla\phi_j\big|\fal{\rho}_s^{II}\big|\phi_i\bra=0$ (cf. Def. \ref{def:quasikom}), which implies
$\rho_{kkji}=\sum_s\fal{\lambda}_s\bla\varphi_k\big|\fal{\rho}_s^{I}\big|\varphi_k\bra\bla
\phi_j\big|\fal{\rho}_s^{II}\big|\phi_i\bra=0$ and, of course, (\ref{tw2:ii}) holds. We have shown that (i) implies
0(\ref{tw2:ii}). By Proposition \ref{tw2} we have $\korelsepf$ and thus, by equivalence of the decompositions of $\rho$,
we get $\korelsep$ which ends the proof. \finito

\vspace{4mm}

The implication in Proposition \ref{tw3} can be partially inverted. We have
\begin{stw}
Let $\rosep$, $\fr$, $\gr$, $dim\hj=n$, $dim\hd=m$. Assume that there exists decomposition $\rosepf$ such that
$\{\fal{\rho}_j^I\}$ is abelian.
Then the following conditions are equivalent: \\
(i) For some decomposition $\roseph$ we have:
\beqar
\{\widehat{\rho}_j^I\}\;\;\textrm{is abelian and}\;\;\{\widehat{\rho}_j^{II}\}\;\;\textrm{is K-quasi-abelian $(K\leq m$)}
\nonumber
\eeqar
(ii) $\korelsep$ \label{tw4}
\end{stw}
\dw $((i)\Rightarrow (ii))$ This implication follows from Proposition \ref{tw3}. \\
$((ii)\Rightarrow (i))$ Let $\{\varphi_i\}$ be the orthonormal basis of $\hj$, in which $\fal{\rho}_j^I$ are diagonal
and $\{\phi_j\}$ be the orthonormal basis consisting of eigenvectors of the reduced density matrix
$\fal{\rho}=Tr_1(\rho)$. Then there exists decomposition
$\rho=\sum_j\widehat{\lambda}_j\widehat{\rho}_j^I\oti\widehat{\rho}_j^{II}$ such that
$\widehat{\rho}_j^I=\big|\varphi_j\bra\bla\varphi_j\big|$ (the family $\{\widehat{\rho}_j^I\}$ is abelian). The matrix
elements of $\rho$ are $\rho_{kluw}=\sum_i\lambda_i\eta_{ikl}\xi_{iuw}$ with
$\eta_{ikl}=\la\varphi_k|\widehat{\rho}_i^I|\varphi_l\ra=\delta_{ik}\delta_{il}$ and
$\xi_{iuw}=\la\phi_u|\widehat{\rho}_i^I|\phi_w\ra$. Obviously, $\korelsep$ implies $\korelseph$. By Proposition
\ref{tw2} it means that
\beq
\rho_{lkji}=\sum_{p=1}^n\sum_{q=1}^m\rho\prw_{pkjq}\rho\prw_{lpqi}\sqrt{\frac{\fal{\rho}_{jj}}{\fal{\rho}_{ii}}}
\label{lem2:2}
\eeq
where $\fal{\rho}_{jj}$ is eigenvalue of $\fal{\rho}$ corresponding to $\phi_j$. Since $\rho_{klji}=0$ if $l\neq k$ (which
implies $\rho\prw_{klji}=0$ for $l\neq k$), (\ref{lem2:2}) reduces to
$\rho_{kkji}=\sum_{q=1}^m\rho\prw_{kkjq}\rho\prw_{kkqi}\sqrt{\fal{\rho}_{jj}/\fal{\rho}_{ii}}$, and
$\sum_q\rho\prw_{kkjq}\rho\prw_{kkqi}=\rho_{kkji}$. Hence, if (\ref{lem2:2}) holds, then for every $j,i\in\{1,\ldots,m\}$
such that $j\neq i$ we have: if $\fal{\rho}_{jj}\neq\fal{\rho}_{ii}$ then $\rho_{kkji}=0$ for $k=1,\ldots,n$. But
$\rho_{kkji}=0$ means $\sum_l\lambda_l\eta_{lkk}\xi_{lji}=0$. Since $\eta_{ikl}=\delta_{ik}\delta_{il}$, we have
$0=\sum_l\lambda_l\eta_{lkk}\xi_{lji}=\sum_l\lambda_l\delta_{lk}\xi_{lji}=\lambda_k\xi_{kji} \;\Rightarrow\;\xi_{kji}=0$ as
from the definition of decomposition $\forall_l\,\lambda_l>0$. Thus, the set $\{\widehat{\rho}_i^{II}\}$ has the following
property: if $\fal{\rho}_{jj}\neq\fal{\rho}_{ii}$ then for every $k\in\{1,\ldots,n\}$ we have
$(\widehat{\rho}_k^{II})_{ji}\equiv\xi_{kji}=0$. From Definition \ref{def:quasikom} it follows that the set
$\{\rho_i^{II}\}$ is K-quasi-abelian for some $K\leq m$. \finito

\subsection{Factorization and nondegeneracy of density matrix} \label{sekcja5c}

The next result provides a criterion for the factorization of correlation functions under the nondegeneracy condition
specified below. To show this equivalence we need the following result:
\begin{lem}
Let $\rosep$, $dim\hj=n$, $dim\hd=m$ and the matrix elements $\rho_{klji}$ satisfy in some product basis $\bazajd$ the
following condition $\rho_{klji}=0$ whenever $k\neq l$ ($\rho_{klji}=0$ whenever $j\neq i$). Then there exists
decomposition $\rosepf$ such that $\{\fal{\rho}_j^I\}$ is abelian ($\{\fal{\rho}_j^{II}\}$ is abelian). \label{lemat1}
\end{lem}
\dw Let us consider the matrix representation of $\rho$, i.e. $\rho=\big[\rho_{klji}\big]_{k,l=1,\ldots,n;\,j,i=1,\ldots,m}$
(cf. (\ref{E*3})). Suppose that $\rho_{klji}=0$ whenever $k\neq l$. Let $\fal{\lambda}_s:=\sum_{p=1}^m\rho_{sspp}$.
Consider the following decomposition of $\rho$:
\beq
\rho=\sum_{s=1}^n\fal{\lambda}_s\fal{\rho}_s^I\oti\fal{\rho}_s^{II} \label{przedst:xx}
\eeq
where
\begin{displaymath}
\fal{\rho}_s^I=\big|\varphi_s\bra\bla\varphi_s\big|\qquad\qquad \big[\fal{\rho}_s^{II}\big]_{ji}= \left\{\begin{array}{ll}
\displaystyle{\frac{1}{\fal{\lambda}_s}\cdot\big[\rho_{ssji}\big]_{j,i=1}^m} &
\textrm{ if }\fal{\lambda}_s>0  \\
0 & \textrm{ if }\fal{\lambda}_s=0
\end{array}\right.\end{displaymath}
One can easily check that (\ref{przedst:xx}) is a well defined decomposition of $\rho$ (in particular if $\fal{\lambda}_s=0$
then $\rho_{ssji}=0$ for $j,i=1,\cdots,m$). Of course $\{\fal{\rho}_s^I\}$ is abelian. The proof of the second statement is
similar. \finito

\vspace{3mm}

Now we are in position to give the promised result which shows a relation between factorization of correlation function
and the spectral properties (nondegeneracy) of density matrix.
\begin{stw}
Let $\rosep$, $\fr$, $\gr$, $dim\hj=n$, $dim\hd=m$. Assume that the reduced density matrix $\rozred(\equiv\trjr)$ has
nondegenerated eigenvalues. Then the following conditions are equivalent:
\begin{displaymath} \begin{array}{l}
(i)\textrm{ there exists decomposition }\displaystyle{\rosepf}\textrm{ such that $\{\fal{\rho}_j^{II}\}$ is abelian} \\
(ii)\textrm{ correlation function can be factorized, i.e. }\korelsep
\end{array}
\end{displaymath} \label{tw5}
\end{stw}
\dw $((i)\Rightarrow (ii))$ This implication follows from Proposition \ref{tw3}. \\
$((ii)\Rightarrow (i))$ Let $\bazaj\si$ be arbitrary orthonormal basis of $\hj$ and $\bazad\sj$ be orthonormal basis of
$\hd$ consisting of eigenvectors of $\fal{\rho}$. If the correlation function factorizes, then from Proposition \ref{tw2} we
have $\rho_{lkji}=\sum_{p=1}^n\sum_{q=1}^m\rho\prw_{pkjq}\rho\prw_{lpqi}\sqrt{\fal{\rho}_{jj}/\fal{\rho}_{ii}}$ with
$\fal{\rho}_{ii}$ - the eigenvalue of $\fal{\rho}$ corresponding to $\phi_i$. This equality and self-adjointness of $\rho$
imply:
\beqar
\sum_{pq}\rho\prw_{pkjq}\rho\prw_{lpqi}\sqrt{\frac{\fal{\rho}_{jj}}{\fal{\rho}_{ii}}}=
\ovl{\sum_{rs}\rho\prw_{rlis}\rho\prw_{krsj}\sqrt{\frac{\fal{\rho}_{ii}}{\fal{\rho}_{jj}}}}=
\sum_{rs}\ovl{\rho\prw_{rlis}}\;\ovl{\rho\prw_{krsj}}\sqrt{\frac{\fal{\rho}_{ii}}{\fal{\rho}_{jj}}}
=\sum_{rs}\rho\prw_{rkjs}\rho\prw_{lrsi}\sqrt{\frac{\fal{\rho}_{ii}}{\fal{\rho}_{jj}}} \nonumber
\eeqar
Multiplying the above equality by $\bigg(\displaystyle{\frac{\fal{\rho}_{jj}}{\fal{\rho}_{ii}}}\bigg)$ gives:
\beqar
\underbrace{\sum_{pq}\rho\prw_{pkjq}\rho\prw_{lpqi}\sqrt{\frac{\fal{\rho}_{jj}}{\fal{\rho}_{ii}}}}
_{\equiv\rho_{lkji}}\cdot\frac{\fal{\rho}_{jj}}{\fal{\rho}_{ii}}=
\underbrace{\sum_{rs}\rho\prw_{rkjs}\rho\prw_{lrsi}\sqrt{\frac{\fal{\rho}_{jj}}{\fal{\rho}_{ii}}}}_{\equiv\rho_{lkji}}
\;\Rightarrow\;\rho_{lkji}\bigg(\frac{\fal{\rho}_{jj}}{\fal{\rho}_{ii}}\bigg)=\rho_{lkji}
\;\Rightarrow\;\rho_{lkji}\big(\fal{\rho}_{jj}-\fal{\rho}_{ii}\big)=0 \nonumber
\eeqar
By the assumption $\fal{\rho}_{jj}\neq\fal{\rho}_{ii}$, hence $\rho_{lkji}=0$ whenever $j\neq i$. Our Proposition follows
then from Lemma \ref{lemat1}. \finito

\vspace{3mm}

The results of this section provide a natural and intrinsic characterization of the two-points correlation function for
block spin-flip dynamics and for the initial separable state. But one question still unanswered is whether the
factorization or non-factorization of such functions is a genuine property for the considered dynamics. To answer this
question we want to show that there are a lot of separable density matrices for which the correlation function $\korel$
can not be factorized. Namely, we have the following:
\begin{stw}
The set of density matrices such that the equality $\korelsep$ does not hold, is dense in $\fal{S}_{sep}$ where
$\fal{S}_{sep}=\{\rho\in S_{sep}:\;Tr_1{\rho}\;\textrm{is invertible}\}$. \label{tw6} \finito
\end{stw}
The proof of Proposition \ref{tw6} is given in Appendix B. \\
We want to complete this section with the observation that there exists a strict connection between the problem of
factorization of the correlation function and the separability of the square root $\rho\prw$. Namely:
\begin{stw}
Let $\rosep$. If there exists decomposition $\rosepfo$ such that one of the following conditions is satisfied:
\begin{displaymath} \begin{array}{l}
(i) \quad \{\fal{\rho}_j^{I}\} \textrm{ is abelian } \\
(ii) \quad \{\fal{\rho}_j^{II}\} \textrm{ is abelian }
\end{array} \end{displaymath}
then $\rho\prw$ is separable. \label{tw9} \finito
\end{stw}
\dw Suppose that there exists decomposition $\rosepfo$ such that $\{\fal{\rho}_j^{I}\}$ is abelian (the proof for the case
when (ii) holds is similar). Let $\{\varphi_i\}$ be the orthonormal basis of $\hj$, in which $\fal{\rho}_j^I$ are diagonal.
Then, there exists decomposition $\rho=\sum_k\widehat{\lambda}_k\widehat{\rho}_k^I\oti\widehat{\rho}_k^{II}$ such that
$\widehat{\rho}_k^I=\big|\varphi_k\bra\bla\varphi_k\big|$. Define matrix $\bar{\rho}$ as follows:
\beqar
\bar{\rho}:=\sum_k\widehat{\lambda}_k\prw\widehat{\rho}_k^I \oti\big(\widehat{\rho}_k^{II}\big)\prw
\nonumber
\eeqar
Note that $\bar{\rho}$ is a linear combination with positive coefficients and matrices $\widehat{\rho}_k^I$ and
$\widehat{\rho}_k^{II}$ are positive operators. To complete the proof we must show that $\bar{\rho}$ is the square root of
$\rho$. We have:
\beqar
\bar{\rho}\cdot\bar{\rho} & = & \bigg(\sum_k\widehat{\lambda}_k\prw\widehat{\rho}_k^I
\oti\big(\widehat{\rho}_k^{II}\big)\prw\bigg)\cdot\bigg(\sum_l\widehat{\lambda}_l\prw\widehat{\rho}_l^I
\oti\big(\widehat{\rho}_l^{II}\big)\prw\bigg)=
\sum_{kl}\widehat{\lambda}_k\prw\widehat{\lambda}_l\prw\cdot\widehat{\rho}_k^I\widehat{\rho}_l^I\oti
\big(\widehat{\rho}_k^{II}\big)\prw\big(\widehat{\rho}_l^{II}\big)\prw \nonu
\eeqar
\beqar
=\sum_{kl}\widehat{\lambda}_k\prw\widehat{\lambda}_l\prw\cdot\widehat{\rho}_k^I\delta_{kl}\oti
\big(\widehat{\rho}_k^{II}\big)\prw\big(\widehat{\rho}_l^{II}\big)\prw
=\sum_{k}\widehat{\lambda}_k\widehat{\rho}_k^I\oti\widehat{\rho}_k^{II}=\rho \nonumber
\eeqar
\finito Note that the above sufficient conditions for the separability of the square root of $\rho$ are essentially weaker
than those for factorization of the correlation function. \\

\section{Conclusions}

Our analysis yields new information about the nature of quantum spin-flip dynamics. It was shown that this type of
evolution can lead to entangled states, so to the family of states encoding quantum correlations. Furthermore, the
detailed analysis of factorization property of the Ursell functions clearly shows that this is an expected phenomenon. It
would be desirable to have the full description of evolution of entanglement but we have not been able to do this. The
main difficulty in carrying out such a description is that we do not know the general characterization of interactions
causing the spin-flip operation. However, we were able to give a detailed analysis of factorization of two-point
correlation functions. We should also emphasize that in our case there is no point in considering such notions as
decoherence. Although the region $\Lambda_I$ over which we perform a local operation (e.g. a block-spin flip) can be
macroscopic, we do not deal with the collective observables. This means that studying the evolution of entanglement as
well as the correlation functions is the proper tool in investigating genuine microscopic properties of the considered
quantum dynamics. These two approaches, studied in Section III and IV, respectively, are not equivalent. It is well known
that if at least one of the two subsystems is a classical system (i.e. the underlying algebra of operators is
commutative), there is no entanglement, even if the second subsystem is a purely quantum one \cite{N4}. On the other hand,
studying the correlation functions allows us to "detect" quantum properties of the block-spin flip dynamics also when
either of the two subsystems is a classical one (cf. Proposition \ref{tw4}). Thus, in that aspect, the correlations based
approach is a more "subtle" tool as far as we examine the problem of the considered dynamics being the genuine quantum
map or not. However, even the detailed analysis of the second Ursell function (the analysis of the third, fourth, ...
Ursell functions is a very difficult task) can not supply enough information about the time evolution of states of the
system. From this point of view the approach given in Section III is much more fruitful.

Turning back to the block-spin flip dynamics, we pointed out that its Markov generator should encode coupling between
the region $\Lambda_I$ and its environment $\Lambda_{II}$. Indeed, the results of Section III and IV say that the
considered dynamics leads to enhancement of correlations. This means that the effect caused by the block-spin flip
operation is strong, and it leads to coupling between two subsystems, therefore to nontrivial interactions. This
enables us to interpret our result as another evidence that $L_p$-approach to quantum dynamics is working well in the
sense that it leads to a fruitful recipe for explicit construction of interesting genuine quantum counterparts of
classical dynamical maps.

Finally, we would like to remark that our results have been obtained for a low-dimensional model. Therefore, the
expected and described properties of block-spin flip type dynamics follow exclusively from the noncommutativity of the
underlying algebra of operators and have nothing to do with any transition from a finite to an infinite model via
thermodynamic limit. Furthermore, we would like to emphasize that the presented theory has a fairly straightforward
generalization to the infinite dimensional case as well as to other quantum jump processes.

{\bf Acknowledgement:} It is a pleasure to thank M. Marciniak for inspiring discussions. The work of (W.A.M) has been
supported by KBN grant PB/0273/PO/99/16.

\vskip 1cm
\noindent{\large\bf Appendix A}
\stepcounter{section}
\vskip 1cm

%\subsection{} \label{app0}
For the convenience of the reader we recall the definition of separable and entangled states in the general setting of
Hilbert spaces. The density matrix $\rho$ (state) on the Hilbert space $\hjd$ is called separable if it can be written or
approximated (in the norm) by the density matrices (states) of the form:
\beqar
\rho=\sum_ip_i\rho_i^1\otimes\rho_i^2,\quad \left(\omega(\cdot)=\sum_ip_i(\omega_i^1\otimes\omega_i^2)(\cdot)\right)
\nonumber
\eeqar
where $p_i\geq 0$, $\sum_ip_i=1$, $\rho_i^{\alpha}$ are density matrices on $\cH_{\alpha}$, $\alpha=1,2$, and
$(\omega_i^1\otimes\omega_i^2)(A\otimes
B)\equiv\omega_i^1(A)\cdot\omega_i^2(B)\equiv(Tr\rho_i^1A)\cdot(Tr\rho_i^2B)\equiv Tr\{\rho_i^1\otimes\rho_i^2\cdot
A\otimes B\}$.
\begin{deff}
Let $\rho$ be the separable state on $\hjd$. We say that every finite sum of the form
$\rho=\sum_i\lambda_i\rho_i^1\otimes\rho_i^2$ is a {decomposition} of the state $\rho$ iff
\beqar
(i) & \qquad\sum_i\lambda_i=1\qquad\forall_i\;\lambda_i>0 \nonumber \\
(ii) & \sum_i\lambda_i\rho_i^1\otimes\rho_i^2=\rho \nonumber
\eeqar
\finito
\end{deff}
The state which is not separable is called non-separable. Denote by $S$ the set of all states on $\hjd$.
\begin{deff}
Non-separable states are called entangled states. The set of entangled states is defined by
\beqar
S_{entangled}\equiv S\backslash S_{sep} \nonumber
\eeqar
where $S_{sep}$ stands for the set of separable states.
\end{deff}
For a discussion of physical aspects of that concept see \cite{R12}, \cite{R13}, \cite{R14}.

\vskip 1cm
\noindent{\large\bf Appendix B}
\stepcounter{section}

\vskip 1cm
%\section{} \label{app1}
In this appendix we give the proof of Proposition \ref{tw6}. Let us introduce the following notation:
$S_{nd}\subset\fal{S}_{sep}$, $\rho\in S_{nd}$ if and only if the eigenvalues of $\fal{\rho}=Tr_1(\rho)$ are not
degenerated, $S_{nf}\subset\fal{S}_{sep}$, $\rho\in S_{nf}$ if and only if $\korel$ can not be factorized,
$S_{ndf}:=S_{nd}\cap S_{nf}$.
\begin{lem}
Let $\;dim\hj,dim\hd<\infty$. Then the set $S_{nd}$ is dense in $\fal{S}_{sep}$ in uniform topology (equivalently, it is
dense in any operator topology). \label{lematgest1}
\end{lem}
\dw Let $\rho\in\fal{S}_{sep}$ and $\rosepn$ be some decomposition of $\rho$. Let $\bazad\sj$ be the orthonormal basis of
$\hd$ such that $\fal{\rho}=\trjr$ is diagonal. Denote by $e_i$ the eigenvalue of $\fal{\rho}$ corresponding to eigenvector
$\phi_i$. Of course, we have $e_i:=\bla\phi_i\big|\fal{\rho}\big|\phi_i\bra$. Without loss of generality we can assume that
only one eigenvalue is degenerated. In particular, we can assume that $e_1=e_2$. Let $\epsilon>0$. We will show that there
exists $\widehat{\rho}\in\Snd$ such that $||\rho-\widehat{\rho}||<\epsilon$. Take $\eta$ such that
$0<\eta<\frac{1}{2}min\big\{\epsilon,\big|e_3-e_1\big|,\ldots,\big|e_m-e_1\big|\big\}$. Define:
\beqar
\forall_{i=1,\ldots,N}\qquad\widehat{\lambda}_i:=\lambda_i\big(1-\eta\big),\quad\widehat{\rho}_i^I:=\rho_i^I,\quad
\widehat{\rho}_i^{II}:=\rho_i^{II} \nonumber
\eeqar
\beqar
\widehat{\lambda}_{N+1}:=\eta,\quad\widehat{\rho}_{N+1}^I:=\frac{1}{dim\hj}id_{\hj},\quad
\widehat{\rho}_{N+1}^{II}:=\big|\phi_1\bra\bla\phi_1\big| \nonumber
\eeqar
and
\beqar
\widehat{\rho}:=\sum_{i=1}^{N+1}\widehat{\lambda}_i\widehat{\rho}_i^I\oti\widehat{\rho}_i^{II} \nonumber
\eeqar
Note that $\widehat{\rho}$ is a well defined density matrix on $\hjd$. Moreover, the reduced density matrix
$Tr_1(\widehat{\rho})=\sum_{i=1}^{N+1}\widehat{\lambda}_i\widehat{\rho}_i^{II}=\fal{\rho}\big(1-\eta\big)+
\eta\big|\phi_1\bra\bla\phi_1\big|$ has only nondegenerated eigenvalues $\widehat{e}_1=e_1$,
$\widehat{e}_2=e_2\big(1-\eta),\ldots,\widehat{e}_m=e_m\big(1-\eta)$, so $\widehat{\rho}\in S_{nd}$. The lack of degeneracy
stems from the choice of $\eta$ because for all $i=3,\ldots,m$ we have $\big|e_i-e_1\big|>2\eta$ and, evidently,
$\big|e_i\eta\big|\leq\eta$. Now, suppose that $\widehat{e}_1=\widehat{e}_j$ for some $j\in\{3,\ldots,m\}$. We have:
\beqar
\widehat{e}_1=\widehat{e}_j & \quad\Leftrightarrow\quad & e_1=e_j\big(1-\eta\big)\quad\Leftrightarrow\quad e_1-e_j=-e_j\eta
\quad\Rightarrow\quad \nonumber \\
& \quad\Rightarrow\quad & \big|e_1-e_j\big|=\big|e_j\eta\big|\quad\Rightarrow\quad 2\eta<\eta
\quad\Leftrightarrow\quad \eta<0 \nonumber
\eeqar
which yields a contradiction, since $\eta$ was assumed to be positive.\\
To complete the proof we must check that the inequality $||\rho-\widehat{\rho}||<\epsilon$ holds. Indeed:
\beqar
||\rho-\widehat{\rho}|| & = & \Big|\Big|\sum_{i=1}^N\lambda_i\rho_i^I\oti\rho_i^{II}-
\sum_{i=1}^{N+1}\widehat{\lambda}_i\widehat{\rho}_i^I\oti\widehat{\rho}_i^{II}\Big|\Big| \nonumber \\
& = & \Big|\Big|\sum_{i=1}^N\lambda_i\rho_i^I\oti\rho_i^{II}- \sum_{i=1}^N\big(1-\eta)\lambda_i\rho_i^I\oti\rho_i^{II}-
\eta\,\widehat{\rho}_{N+1}^I\oti\widehat{\rho}_{N+1}^{II} \Big|\Big| \nonumber \\
& = & \Big|\Big|\eta\sum_{i=1}^N\lambda_i\rho_i^I\oti\rho_i^{II}-
\eta\,\widehat{\rho}_{N+1}^I\oti\widehat{\rho}_{N+1}^{II} \Big|\Big| \nonumber \\
& = & \eta\big|\big|\rho-\widehat{\rho}_{N+1}^I\oti\widehat{\rho}_{N+1}^{II}\big|\big| \leq
\eta\big(||\rho||+||\widehat{\rho}_{N+1}^I\oti\widehat{\rho}_{N+1}^{II}||\big) \leq 2\eta < \epsilon \nonumber
\eeqar
\finito
\begin{lem}
Let $\;dim\hj,dim\hd<\infty$. Then the set $S_{ndf}$ is dense in $S_{nd}$ in uniform topology (equivalently, it is dense in
any operator topology). \label{lematgest2}
\end{lem}
\dw Let $\rho\in S_{nd}$ and $dim\hd=m$. Suppose that $\korel$ can be factorized.
Then, from the relation between factorization and nondegeneracy of density matrix (see section \ref{sekcja5c}) there
exists decomposition $\rosepn$ such that $\{\rho_i^{II}\}$ is abelian. We can assume that $N$ equals $dim\hd$ and
$\rho_i^{II}=\big|\phi_i\bra\bla\phi_i\big|$, where $\bazad$ is the orthonormal basis of $\hd$ such that $\fal{\rho}=\trjr$
is abelian. Denote by $e_i$ the eigenvalue of $\fal{\rho}$ corresponding to the eigenvector $\phi_i$ (we have
$e_i=\bla\phi_i\big|\fal{\rho}\big|\phi_i\bra$). Without loss of generality we can assume that eigenvalues of
$\fal{\rho}$ are ordered decreasingly, i.e. $e_1>e_2>\ldots\;$. \\
Let $\epsilon>0$. We will show that there exists $\widehat{\rho}\in S_{ndf}$ such that $||\rho-\widehat{\rho}||<\epsilon$.
Take $\eta$ such that $0<\eta<\epsilon\slash 2$. Let $\bazaj\si$ be arbitrary but fixed orthonormal basis of $\hj$. Define:
\beqar
\forall_{i=1,\ldots,N}\qquad\widehat{\lambda}_i:=\lambda_i\big(1-\eta\big),\quad\widehat{\rho}_i^I:=\rho_i^I,\quad
\widehat{\rho}_i^{II}:=\rho_i^{II} \nonumber
\eeqar
\beqar
\widehat{\lambda}_{N+1}:=\frac{1}{2}\eta,\quad\widehat{\rho}_{N+1}^I:=\big|\varphi_1\bra\bla\varphi_1\big|,\quad
\widehat{\rho}_{N+1}^{II}:=\frac{1}{2}\big|\varphi_1\bra\bla\varphi_1\big|+\frac{1}{2}\big|\varphi_2\bra\bla\varphi_2\big|
+\frac{i}{4}\big|\varphi_1\bra\bla\varphi_2\big|-\frac{i}{4}\big|\varphi_2\bra\bla\varphi_1\big| \nonu
\eeqar
\beqar
\widehat{\lambda}_{N+2}:=\frac{1}{2}\eta,\quad\widehat{\rho}_{N+2}^I:=\big|\varphi_2\bra\bla\varphi_2\big|,\quad
\widehat{\rho}_{N+1}^{II}:=\frac{1}{2}\big|\varphi_1\bra\bla\varphi_1\big|+\frac{1}{2}\big|\varphi_2\bra\bla\varphi_2\big|
-\frac{i}{4}\big|\varphi_1\bra\bla\varphi_2\big|+\frac{i}{4}\big|\varphi_2\bra\bla\varphi_1\big| \nonu
\eeqar
and
\beqar
\widehat{\rho}:=\sum_{i=1}^{N+2}\widehat{\lambda}_i\widehat{\rho}_i^I\oti\widehat{\rho}_i^{II} \nonu
\eeqar
Note that $\widehat{\rho}$ is well defined density matrix on $\hjd$. Moreover,
$Tr_1(\widehat{\rho})=\sum_{i=1}^{N+2}\widehat{\lambda}_i\widehat{\rho}_i^{II}=\fal{\rho}\big(1-\eta\big)+
\frac{1}{2}\eta\big(\big|\phi_1\bra\bla\phi_1\big|+\big|\phi_2\bra\bla\phi_2\big|\big)$ has only nondegenerated eigenvalues
$\widehat{e}_1=e_1\big(1-\frac{1}{2}\eta\big)$, $\widehat{e}_2=e_2\big(1-\frac{1}{2}\eta\big)$,
$\widehat{e}_3=e_3\big(1-\eta),\ldots,\widehat{e}_m=e_m\big(1-\eta)$, so $\widehat{\rho}\in S_{nd}$. \\
Now we aim at showing that $\widehat{\rho}\in S_{ndf}$. It is enough to show that there is no decomposition
$\widehat{\rho}=\sum_{i=1}^{\breve N}\breve{\lambda}_i\breve{\rho}_i^I\oti\breve{\rho}_i^{II}$ for which
$\{\breve{\rho}_i^{II}\}$ is abelian, since due to the relation between factorization and nondegeneracy of density matrix
(cf. section \ref{sekcja5c}) it is equivalent to the fact that $\bla E(f)g\bra_{\widehat{\rho}}$ does not factorize.
Suppose that there exists such a decomposition with $\{\breve{\rho}_i^{II}\}$ abelian. We can assume that $\breve{N}$
equals $dim\hd$ and $\breve\rho_i^{II}=\big|\phi_i\bra\bla\phi_i\big|$. Then, we have:
\beqar
& & \sum_{i=1}^{\breve N}\breve{\lambda}_i\breve{\rho}_i^I\oti\breve{\rho}_i^{II}=
\sum_{i=1}^{N+2}\widehat{\lambda}_i\widehat{\rho}_i^I\oti\widehat{\rho}_i^{II} \nonumber \\
& \Rightarrow & \sum_{i=1}^m\breve{\lambda}_i\breve{\rho}_i^I\oti\big|\phi_i\bra\bla\phi_i\big|=
\sum_{i=1}^m\widehat{\lambda}_i\widehat{\rho}_i^I\oti\big|\phi_i\bra\bla\phi_i\big|+
\widehat{\lambda}_{N+1}\widehat{\rho}_{N+1}^I\oti\widehat{\rho}_{N+1}^{II}+
\widehat{\lambda}_{N+2}\widehat{\rho}_{N+2}^I\oti\widehat{\rho}_{N+2}^{II} \nonumber \\
& \Rightarrow & \sum_{i=1}^m\big(\breve{\lambda}_i\breve{\rho}_i^I-\widehat{\lambda}_i\widehat{\rho}_i^I\big)
\oti\big|\phi_i\bra\bla\phi_i\big|= \widehat{\lambda}_{N+1}\widehat{\rho}_{N+1}^I\oti\widehat{\rho}_{N+1}^{II}+
\widehat{\lambda}_{N+2}\widehat{\rho}_{N+2}^I\oti\widehat{\rho}_{N+2}^{II} \nonumber
\eeqar
Since $(\breve{\lambda}_i\breve{\rho}_i^I-\widehat{\lambda}_i\widehat{\rho}_i^I) \oti\big|\phi_i\bra\bla\phi_i\big|$ are
linearly independent, we have $\breve{\lambda}_i\breve{\rho}_i^I=\widehat{\lambda}_i\widehat{\rho}_i^I$ for $i=3,\ldots,m$,
which leads to the following equality:
\beqar
\sum_{i=1}^2\big(\breve{\lambda}_i\breve{\rho}_i^I-
\widehat{\lambda}_i\widehat{\rho}_i^I\big)\oti\big|\phi_i\bra\bla\phi_i\big|=
\widehat{\lambda}_{N+1}\widehat{\rho}_{N+1}^I\oti\widehat{\rho}_{N+1}^{II}+
\widehat{\lambda}_{N+2}\widehat{\rho}_{N+2}^I\oti\widehat{\rho}_{N+2}^{II} \nonumber
\eeqar
Denote the left-hand side and the right-hand side of the above equality by $L$ i $P$, respectively, we have:
\beqar
\bla\varphi_1\oti\phi_1\big|L\big|\varphi_1\oti\phi_2\bra=0\quad\textrm{and}\quad
\bla\varphi_1\oti\phi_1\big|P\big|\varphi_1\oti\phi_2\bra=\frac{i}{8}\eta\neq 0 \nonumber
\eeqar
which yields a contradiction. Thus, $\widehat{\rho}\in S_{ndf}$. \\
To complete the proof we must check that the inequality $||\rho-\widehat{\rho}||<\epsilon$ holds. Indeed:
\beqar
||\rho-\widehat{\rho}|| & = & \Big|\Big|\sum_{i=1}^N\lambda_i\rho_i^I\oti\rho_i^{II}-
\sum_{i=1}^{N+2}\widehat{\lambda}_i\widehat{\rho}_i^I\oti\widehat{\rho}_i^{II}\Big|\Big| \nonumber \\
& = & \Big|\Big|\sum_{i=1}^N\lambda_i\rho_i^I\oti\rho_i^{II}- \sum_{i=1}^N\big(1-\eta)\lambda_i\rho_i^I\oti\rho_i^{II}-
\frac{1}{2}\eta\,\widehat{\rho}_{N+1}^I\oti\widehat{\rho}_{N+1}^{II}-
\frac{1}{2}\eta\,\widehat{\rho}_{N+2}^I\oti\widehat{\rho}_{N+2}^{II} \Big|\Big| \nonumber \\
& = & \Big|\Big|\eta\sum_{i=1}^N\lambda_i\rho_i^I\oti\rho_i^{II}-
\frac{1}{2}\eta\,\widehat{\rho}_{N+1}^I\oti\widehat{\rho}_{N+1}^{II}-
\frac{1}{2}\eta\,\widehat{\rho}_{N+2}^I\oti\widehat{\rho}_{N+2}^{II} \Big|\Big| \nonumber \\
& = & \eta\big|\big|\rho-\frac{1}{2}\widehat{\rho}_{N+1}^I\oti\widehat{\rho}_{N+1}^{II}
-\frac{1}{2}\widehat{\rho}_{N+2}^I\oti\widehat{\rho}_{N+2}^{II}\big|\big| \nonumber \\
& \leq & \eta\big(||\rho||+\frac{1}{2}||\widehat{\rho}_{N+1}^I\oti\widehat{\rho}_{N+1}^{II}||
+\frac{1}{2}||\widehat{\rho}_{N+2}^I\oti\widehat{\rho}_{N+2}^{II}||\big) \leq 2\eta < \epsilon \nonumber
\eeqar
\finito \emph{Proof (of Proposition \ref{tw6}).} The following inclusions hold: $S_{ndf}\subset S_{nd}\subset\fal{S}_{sep}$.
According to Lemma \ref{lematgest1} and Lemma \ref{lematgest2}, $S_{nd}$ is dense in $\fal{S}_{sep}$ and $S_{ndf}$ is dense
in $S_{nd}$, respectively. It means that $S_{ndf}$ is dense in $\fal{S}_{sep}$. Moreover, we have: $S_{ndf}\subset
S_{nf}\subset\fal{S}_{sep}$ which implies that $S_{nf}$ is dense in $\fal{S}_{sep}$. The proof is completed. \finito

\end{document}